# Switching Portfolios

**Yoram Singer**
AT&T Labs, Room A277
180 Park Avenue, Florham Park, NJ 07932
singer@research.att.com

## Abstract

A constant rebalanced portfolio is an asset alloca-
tion algorithm which keeps the same distribution
of wealth among a set of assets along a period of
time. Recently, there has been work on on-line
portfolio selection algorithms which are compet-
itive with the *best* constant rebalanced portfolio
determined in *hindsight* [6, 11, 8]. By their na-
ture, these algorithms employ the assumption that
high returns can be achieved using a fixed asset
allocation strategy. However, stock markets are
far from being stationary and in many cases the
wealth achieved by a constant rebalanced portfo-
lio is much smaller than the wealth achieved by an
ad-hoc investment strategy that adapts to changes
in the market. In this paper we present an efficient
Bayesian portfolio selection algorithm that is able
to track a changing market. We also describe a
simple extension of the algorithm for the case
of a general transaction cost, including the trans-
actions cost models recently investigated in [4].
We provide a simple analysis of the competitive-
ness of the algorithm and check its performance
on real stock data from the New York Stock Ex-
change accumulated during a 22-year period.

## 1 Introduction

In this work we represent a stock market as a vector of
stocks denoted by $\mathbf{x} = (x_1, x_2, \ldots, x_N)$ where $N$ is the
number of stocks and $x_i$ is the price relative change. That
is, $x_i$ is the ratio of the next day's opening price of the $i$th
stock to its opening price on the current day. A portfolio is
defined by a weight vector $\mathbf{w} = (w_1, w_2, \ldots, w_N)$ such that
$w_i \geq 0$ and $\sum_{i=1}^{N} w_i = 1$. The $i$th entry of a portfolio $\mathbf{w}$ is
the proportion of the total portfolio value invested in the $i$th
asset. Therefore, if the total wealth is $S$ and the worth of
the $i$th asset is $S^i$ then $w_i = S^i/S$. Given a portfolio $\mathbf{w}$ and
the vector of price relatives $\mathbf{x}$, investors using this portfolio
increase (or decrease) their wealth from one morning to the
next by a factor of $S = \mathbf{w} \cdot \mathbf{x} = \sum_{i=1}^{N} w_i x_i$.

Naturally, we wish to maximize $S$. However, $S$ is
a random variable, therefore, there is a debate over the
choice of a good distribution for $S$. The standard theory of
stock market investment is based on methods that consider
the first (mean) and the second (variance) moments of $S$.
Typically, the objective is to maximize the *expected* value of
$S$, subject to constraints on the variance, which corresponds
to the risk induced by using the portfolio $\mathbf{w}$. Calculating
the mean and the variance is an easier task than dealing with
the entire distribution of $S$. The mean-variance approach is
basis of the Sharpe-Markowitz [15] theory of investments
in the stock market which is used by business analysts. As
mentioned above, the first moments of the random variable
$S$ gives us information on the *expected* long time behavior
under i.i.d assumptions of the prices relatives. However, in
stock markets one normally reinvests every day so that the
total wealth achieved is a product of the individual wealths
achieved on each day. Furthermore, future behavior of
stock markets is not *independent* of the past. For the above
reasons, Bayesian distributional methods (see for instance
[13, 2, 5, 7, 1]), that use adaptive investment strategies for
rebalanced portfolios, have been developed.

The following example (taken from [11]) demonstrates
the power of rebalancing. Assume that two investments
are available. The first is a risk-free, no-growth investment
stock whose value never changes. The second investment is
a hypothetical highly volatile stock. On even days, the value
of this stock doubles and on odd days its value is halved.
The relative returns of the first stock can be described by
the sequence $1, 1, 1, \ldots$ and of the second by the sequence
$\frac{1}{2}, 2, \frac{1}{2}, 2, \ldots$. Neither investment alone can increase in
value by more than a factor of 2, but a strategy combining
the two investments can grow exponentially. One such
strategy splits the investor's total wealth evenly between
the two investments, and maintains this even split at the end
of each day. On odd days the relative wealth decreases by
a factor of $\frac{1}{2} \times 1 + \frac{1}{2} \times \frac{1}{2} = \frac{3}{4}$. However, on even days the
relative wealth grows by $\frac{1}{2} \times 1 + \frac{1}{2} \times 2 = \frac{3}{2}$. Thus, after two
consecutive trading days the investor's wealth grows by a
factor of $\frac{3}{4} \times \frac{3}{2} = \frac{9}{8}$. It takes only twelve days to double
the wealth, and over $2n$ trading days the wealth grows by a
factor of $\left(\frac{9}{8}\right)^n$.

Investment strategies of the type above are called con-
stant rebalanced portfolios (CRP). Formally, a CRP is an
asset allocation algorithm which keeps the same distribu-
tion of wealth among a set of assets along a period of time.

Recent work on distributional methods have focused on on-line portfolio selection algorithms which are competitive with the *best* constant rebalanced portfolio determined in *hindsight* [6, 11, 8]. Most of the portfolio rebalancing algorithms are computationally intractable as the time to update the portfolio vector after each trading days grow exponentially with the number of stocks in the portfolio. Furthermore, these algorithms do not perform well when the market is changing as the following simple example shows.

Assume that two hypothetical highly volatile stocks are available. The value of the first stock increases by a factor of $\frac{3}{2}$ on each of the first $n$ trading periods. Then the stock changes its behavior and the value of the stock falls by a factor of 4 on each of the next $n$ trading periods. The second stock behaves in an opposite manner. Its value falls by a factor of 4 on each of the first $n$ trading periods and increases by a factor of $\frac{3}{2}$ on each of the second $n$ trading days. The relative price change of the first stock can be described by the sequence $\frac{3}{2}, \frac{3}{2}, \ldots, \frac{3}{2}, \frac{1}{4}, \frac{1}{4}, \ldots, \frac{1}{4}$ and of the second by the sequence $\frac{1}{4}, \frac{1}{4}, \ldots, \frac{1}{4}, \frac{3}{2}, \frac{3}{2}, \ldots, \frac{3}{2}$. Investing all of the initial wealth in any of the two stocks results in a fatal loss of almost all of the initial investment. Furthermore, it is easy to verify that the best constant rebalanced portfolio would redistribute the wealth evenly after each trading day, resulting in an exponentially fast wealth decay. Thus, any competitive rebalancing portfolio selection algorithm, such as Cover's universal portfolio algorithm [6], would result in a similar miserable performance. In contrast, a prescient investor who puts all of her money on the first stock for $n$ days and then switches to the second stock would enjoy an enormous profit even in the presence of hefty transaction costs. The switching portfolios algorithm presented in this paper attempts to track the optimal strategy of a prescient investor without any prior knowledge about the actual form of the optimal strategy.

The algorithm proposed here copes with changing markets by considering the possibility that the market changes its behavior after *each* trading day. The algorithm is provided with a set of investment strategies. The investment strategies need not be complex. In this paper we mainly use the simplest strategies one can think of, namely, investing the accumulated wealth in one asset solely. The first version we present assumes that the apriori *duration* of using a given investment strategy is geometrically distributed with known parameters. If, in addition, the set of investment strategies includes only pure strategies, i.e., holding a single asset/stock, then a constant time per asset is required by the algorithm in order to compute wealth redistribution after each trading day. We also give a more general version that does not employ a duration model with fixed parameters. This version achieves better competitiveness bounds but it is more computationally expensive. Both versions can be easily modified to deal with various models of transaction costs. Throughout the paper, we give examples using daily stock market data from the New York Stock Exchange accumulated during a 22-year period and compare the results to the portfolio selection algorithms described in previous works [6, 11, 4], in particular Cover's universal portfolio algorithm and its extension in the presence of transaction costs. On this data, the portfolio selection algorithm we suggest outperforms the universal portfolio algorithm, with and without transaction costs.

The paper is organized as follows. Section 2 provides a short overview of constant rebalanced portfolios. In Section 3 we describe the switching portfolio algorithms, analyze its competitiveness properties, and compare its performance to other portfolio selection algorithms. Section 4 describes a simple extension of the algorithm in the presence of transaction costs. We discuss possible future research directions and conclude in Section 5.

## 2    Rebalanced portfolios

Let us first review again the definitions from the previous section. Consider a portfolio containing $N$ stocks. On each trading day the performance of the stocks can be described by a vector of *price relatives*, denoted by $\mathbf{x} = (x_1, x_2, \ldots, x_N)$ where $x_i$ is the next day's opening price of the $i$th stock divided by its opening price on the current day. Thus, the value of an investment in stock $i$ increases (or falls) by $x_i$ times its previous value from one morning to the next. A portfolio is defined by a weight vector $\mathbf{w} = (w_1, w_2, \ldots, w_N)$ such that $w_i \geq 0$ and $\sum_{i=1}^{N} w_i = 1$. The $i$th entry of a portfolio $\mathbf{w}$ is the proportion of the total portfolio value invested in the $i$th asset. Therefore, if the total wealth is $S$ and the worth of the $i$th asset is $S^i$ then $w_i = S^i/S$. Given a portfolio $\mathbf{w}$ and the vector of price relatives $\mathbf{x}$, investors using this portfolio increase (or decrease) their wealth from one morning to the next by a factor of $\mathbf{w} \cdot \mathbf{x} = \sum_{i=1}^{N} w_i x_i$.

Recent work in on-line portfolio selection algorithms has focused on changing an ensemble of portfolio vectors based on past performance. That is, at the start of each day $t$, the portfolio selection algorithm gets the previous price relatives of the stock market $\mathbf{x}^1, \ldots, \mathbf{x}^{t-1}$. From this information, the algorithm immediately selects its portfolio $\mathbf{w}^t$ for the day. At the beginning of the next day (day $t + 1$), the price relatives for day $t$ are observed and the investor's wealth increases by a factor of $\mathbf{w}^t \cdot \mathbf{x}^t$. Over time, a sequence of price relatives $\mathbf{x}^1, \mathbf{x}^2, \ldots, \mathbf{x}^T$ is observed and a sequence of portfolios $\mathbf{w}^1, \mathbf{w}^2, \ldots, \mathbf{w}^T$ is selected. From the beginning of day 1 through the beginning of day $T + 1$, wealth will have increased by a factor of

$$S_T(\{\mathbf{w}^t\}, \{\mathbf{x}^t\}) \stackrel{\text{def}}{=} \prod_{t=1}^{T} \mathbf{w}^t \cdot \mathbf{x}^t$$

or, alternatively, the logarithm of the increase is

$$LS_T(\{\mathbf{w}^t\}, \{\mathbf{x}^t\}) \stackrel{\text{def}}{=} \sum_{t=1}^{T} \log\left(\mathbf{w}^t \cdot \mathbf{x}^t\right) .$$

Cover (1991) defined a restricted class of investment strategies, called constant rebalanced portfolios (CRP). As noted before, a CRP is rebalanced each day so that a fixed fraction of wealth is held in each of the underlying investments. Therefore, a constant rebalanced portfolio strategy employs the same investment vector $\mathbf{w}$ on each trading day and the resulting wealth and normalized logarithmic wealth after $T$ trading days are

$$S_T(\mathbf{w}) = \prod_{t=1}^{T} \mathbf{w} \cdot \mathbf{x}^t \ , \ \ LS_T(\mathbf{w}) = \sum_{t=1}^{T} \log\left(\mathbf{w} \cdot \mathbf{x}^t\right) \ .$$

Note that such a strategy might require vast amounts of trading, since at the beginning of each trading day the investment proportions are rebalanced back to the vector $\mathbf{w}$. Given a sequence of daily price relatives we can define, in retrospect, the best rebalanced portfolio vector which would have achieved the maximum wealth $S_T$, and hence also the maximum logarithmic wealth, $LS_T$. This portfolio vector is denoted by $\mathbf{w}^\star$. That is,

$$\mathbf{w}^\star \stackrel{\text{def}}{=} \arg\max_{\mathbf{w}} S_T(\mathbf{w}) = \arg\max_{\mathbf{w}} LS_T(\mathbf{w}) \ ,$$

where the maximum is taken over all possible portfolio vectors (i.e., vectors in $\mathbb{R}^N$ with non-negative components that sum to one). Recent work on portfolio selection has focused on on-line weight allocation algorithms that achieve the same asymptotic growth in normalized logarithmic wealth. An on-line algorithm that achieves such an asymptotic behavior is called a *universal* portfolio. Recently, Blum and Kalai (1997) extended the notion of universal portfolios and described a general rebalancing algorithm in the presence of transactions costs. The time complexity of Cover's original universal portfolio algorithm grows exponentially with the size of the portfolio. Hence, it is suitable for portfolios that contain a small number of stocks. Blum and Kalai also suggested an efficient sampling technique for calculating Cover's universal portfolio, with and without transaction costs. Throughout the paper we use Blum and Kalai's implementation of universal portfolios as our straw-man for comparison.

## 3   Switching portfolios

In this section we provide two versions of our portfolio selection algorithm while ignoring transaction costs. A simple modification of the algorithm that takes transaction costs into account is discussed in the next section.

In contrast to previous work which has focused on finding a good portfolio vector, we instead assume that we are given a set of possible investment strategies which we term basic strategies. A basic strategy need not be complex. In fact, with the benefit of hindsight, on each day one can invest all of one's wealth in the single best-performing asset for that day. We thus use pure investment strategies, i.e., strategies that invest all wealth in a single asset, as our basic investment strategies. Clearly, we do not have the luxury of foreseeing future behavior of stock markets. However, as we show, it is possible to track an investment regime that switches from one investment strategy to another as the market changes its behavior. We do so by employing a mixture of all possible switching regimes. This mixture based approach enables us to hedge our bets against the individual switching regimes. We associate a prior probability with each switching regime. The weights are distributed among the different possible switching regimes such that more complicated regimes, that frequently switches from one strategy to another, are *apriori* less favorable. We then let the evidence, i.e., the actual returns, dictate which investment strategy to use.

In summary, our approach to portfolio selection is as follows. We first decide upon a set of investment strategies. We then choose a prior distribution over the possible switching sequences from one investment strategy to another. This prior distribution is recursive in order to enable an efficient evaluation of the portfolio vector. Last, we combine the actual return of each strategy on each day with the prior probability distribution over switching regimes to decide upon a new portfolio vector before each trading day. The wealth achieved by the switching portfolios algorithm is no worse than the wealth achieved by any specific switching regime times the prior probability of that regime. We therefore can give a simple lower bound on the minimal wealth achieved by our algorithm compared to the wealth achieved by *any* of the available switching regimes.

The first version of the switching portfolios algorithm assumes that the duration of using one strategy is geometrically distributed with a given parameter $\gamma$. Thus, if we started using the $i$th investment strategy at time $t_0$, then the *apriori* probability of using this strategy through time $t_1$ (and then switching to a new strategy) is $(1-\gamma)^{t_1-t_0}\gamma$. An investment switching regime $\mathcal{Q}$ for $T$ trading days is described in terms of two lists, $(t_1 t_2 \ldots t_l)$ and $(i_1 i_2 \ldots i_l i_{l+1})$, where the $t_j$'s are indices of the trading days *after* which we switched to a new investment strategy and $i_j$'s are the indices of the basic investment strategies used. Defining $t_0 = 0$ and assuming that a new strategy is picked uniformly at random, the *apriori* probability of using a switching regime $\mathcal{Q}$ from $t = 1$ through $t = T$ is

$$\begin{aligned} P_0(\mathcal{Q}) &= \frac{1}{N}\left(1-\gamma\right)^{T-t_l-1} \\ &\quad \left[\prod_{i=1}^{l}\left(1-\gamma\right)^{t_i-t_{i-1}-1}\gamma\,\frac{1}{N-1}\right] \quad (1) \\ &= \frac{1}{N(N-1)^l}\,\gamma^l\left(1-\gamma\right)^{T-l-1} \ , \quad (2) \end{aligned}$$

where $N$ is the number of investment strategies which in the case of pure strategies is also the number of different assets/stocks. The wealth achieved by a switching regime $\mathcal{Q}$ after $T$ trading days, $S_T(\mathcal{Q})$, is the product the returns of the investment strategies used by the regime, where for pure strategies it is simply

$$S_T(\mathcal{Q}) = \prod_{j=1}^{l+1}\prod_{t=t_{j-1}+1}^{t_j} x_{i_j}^t \ . \quad (3)$$

Therefore, the total accumulated wealth achieved by the mixture of all switching regimes, where each regime is weighted by its prior probability, is simply $\sum_{\mathcal{Q}'} P_0(\mathcal{Q}')S_T(\mathcal{Q}')$. Evaluating this sum directly is clearly infeasible since the number of different switching regimes grows *exponentially* fast. However, since the geometric distribution is memoryless, we can calculate the sum in constant time per asset for each trading day as we now describe.

Let $S_i^t$ be the worth of the $i$th asset after $t$ trading days. Then, at trading day $t+1$, we either stay with the current

(pure) strategy with probability $1 - \gamma$, and therefore keep holding the $i$th asset, or switch to a new strategy (with probability $\gamma$) by redistributing wealth among all other assets. Put another way, the worth of the $i$th asset is the sum of two terms: the first is the previous worth of the asset times the probability that we kept using the $i$th investment strategy on the current trading day; the second is worth of all other assets time the probability that we switched to the $i$th pure investment strategy. More formally, the worth of the $i$th asset after trading day $t+1$ is,

$$S_{t+1}^i = \left( \overbrace{(1-\gamma)S_t^i}^{\text{stay with } i} + \overbrace{\frac{\gamma}{N-1}\sum_{j \neq i} S_t^j}^{\text{switch from } j \text{ to } i} \right) x_i^{t+1} \quad (4)$$

$$= \left( (1 - \tfrac{\gamma N}{N-1})S_t^i + \tfrac{\gamma}{N-1}\sum_{j=1}^N S_t^j \right) x_i^{t+1}. \quad (5)$$

The above equations give a simple procedure for incorporating the prior probability over switching regimes with the actual return. The above scheme can be directly described as a portfolio weight update. Writing $w_t^i = S_t^i / \sum_j S_t^j$, the weight $w_{t+1}^i$ *before* trading day $t+1$ can be described in terms of $w_t^i$ as follows,

$$w_{t+1}^i = \left( 1 - \gamma - \frac{\gamma}{N-1} \right) w_t^i + \frac{\gamma}{N-1}. \quad (6)$$

This portfolio weight update scheme resembles the *fixed-share* weight update used by Herbster and Warmuth (1995) for tracking the best expert in a binary prediction setting. The analysis presented in [12] does not simply carry over to our setting of unbounded returns. Since the next version of the algorithm includes this version as a special case, we defer the analysis of the first version to the end of the section.

We compared the performance of the switching portfolios algorithm to the performance of Cover's universal portfolio algorithm [6], the multiplicative weight update algorithm [11] (denoted by EG($\eta$)), the best constituent stock, and the best constant rebalanced portfolio. We compared the results for all subsets of stocks considered in the experiments described in [6]. The results are summarized in Table 1. In order to achieve high yields in the absence of transactions costs, each asset need to be held only a short time before selling it and buying a new asset. For the NYSE stock data we found that a reasonable time to hold a single stock is typically a few days. To have a fair comparison we set $\gamma = 1/3$ in the all the experiments reported in Table 1, regardless of the stocks constituting the portfolios. In all the experiments reported, the wealths achieved by all the algorithms are calculated assuming an initial investment of one unit before the first trading day. Note that despite the simplicity of the switching portfolio algorithm with a fixed switching probability, it achieves higher yields than the universal portfolio algorithm.

The drawback of using a duration model with a fixed parameter $\gamma$ is that the switching distribution is constrained through this parameter which we need to set in advance. We now give a second version of the switching portfolios algorithm that does not need such prior knowledge. Instead

of setting $\gamma$ to a predefined value, we let $\gamma$ vary in time and define

$$\hat{\gamma}(\Delta t) \stackrel{\text{def}}{=} \frac{1/2}{\Delta t + 1},$$

to be the switching portability after using the same investment strategy for $\Delta t$ consecutive trading days. Therefore, the probability of staying with the strategy increases the more we use the strategy. This parameterization is often used in compression for encoding binary sequences (see for instance [14] and the references therein). Furthermore, we later show that this choice of parameterization for switching probability lets us derive a simple competitiveness bound for the switching portfolios algorithm. Combining this adaptive duration model with the switching portfolios algorithm results in a prior over switching sequences that favors those switching sequences which alternate between the strategies rather infrequently.

Denote by $S_{t,t_0}^i$ the worth of the $i$th asset after the $t$th trading day given that we started holding the asset on trading day $t_0$. We now need to update the worth of each asset based on the start date of the corresponding (pure) investment strategy. The wealth update scheme becomes,

$$S_{t+1,t_0}^i = (1 - \hat{\gamma}(t - t_0)) S_{t,t_0}^i x_i^{t+1}$$
$$= \frac{t - t_0 + 1/2}{t - t_0 + 1} S_{t,t_0}^i x_i^{t+1} \quad (7)$$

$$S_{t+1,t+1}^i = \left( \sum_{j \neq i} \sum_{t_0=1}^t \hat{\gamma}(t - t_0) S_{t,t_0}^j \right) x_i^{t+1}$$
$$= \left( \sum_{j \neq i} \sum_{t_0=1}^t \frac{1}{2(t - t_0 + 1)} S_{t,t_0}^j \right) x_i^{t+1} \quad (8)$$

Note that this more general version has a price. The above wealth update can no longer be computed in a constant time per stock. Since at trading day $t$ we perform $O(t)$ operations, the overall time complexity of the algorithm is $O(t^2)$ (as opposed to $O(t)$ for the first version). Furthermore, there is no equivalent update that directly manipulates a portfolio weight vector. In addition, in the absence of transaction costs, the assumption that there are investment strategies that switch from one asset to another rather *infrequently* and yet yield high returns does hold for the stocks appearing in Table 1. Indeed, when there are no transaction costs, the version with adaptive switching probability achieves yields similar to, and sometime smaller than, the the version with a fixed switching probability. As we discuss in the next section, the usage of the adaptive switching probability scheme becomes useful in the presence of transaction costs. Then, investment strategies that switch rather infrequently are more likely to achieve high returns.

We now discuss the competitiveness properties of the switching portfolios algorithm. We compare the performance of the on-line algorithm to *any* switching regime that can be determined in hindsight. In the following analysis, we use log to denote the base 2 logarithm. To derive a lower bound on wealth achieved by the switching portfolios algorithm we need the following lemma.

| Stocks | Best Stock | BCRP | EG($\eta$) ($\eta = 0.05$) | Universal Portfolio | Switching ($\gamma = 1/3$) |
|---|---|---|---|---|---|
| Iroquois & Kin Ark | 8.92 | 73.70 | 70.85 | 39.97 | 52.55 |
| Com. Met. & Kin Ark | 52.02 | 144.00 | 117.15 | 80.54 | 89.67 |
| Com. Met. & Mei. Corp. | 52.02 | 102.96 | 97.93 | 74.08 | 92.73 |
| IBM & Coca-Cola | 13.36 | 15.07 | 14.90 | 14.24 | 14.96 |

Table 1: Comparison of wealths achieved by the various on-line portfolio selection algorithms. For all the portfolios considered, we give the total wealth achieved by the best constituent stock in the portfolio, the best constant-rebalanced portfolio (BCRP) computed in hindsight from the entire price relatives sequence, the EG($\eta$)-update rule (Helmbold et. al., 1996), Cover's universal portfolio algorithm, and the switching portfolios algorithm with a fixed switching probability.

**Lemma 1**

$$-\log \left( \prod_{i=0}^{n-1} \frac{i+1/2}{i+1} \right) \leq \frac{1}{2}\log(n) + 1 \ .$$

**Proof** The lemma is a special case of a general theorem by Krichevsky and Trofimov (1981). For completeness we now give a simple proof of the lemma. Let

$$g(n) \stackrel{\text{def}}{=} \sqrt{n} \prod_{i=1}^{n} \frac{i-1/2}{i} \ .$$

First note that $g(2) > g(1) = \frac{1}{2}$. We now show for that $n \geq 1$, $g(n+1) > g(n)$. This simply follows from:

$$
\begin{aligned}
g(n+1) &= \sqrt{n+1} \prod_{i=1}^{n+1} \frac{i-1/2}{i} \\
&= \sqrt{n+1} \, \frac{n+1/2}{n+1} \prod_{i=1}^{n} \frac{i-1/2}{i} \\
&= \left( \frac{(n+1/2)^2}{n(n+1)} \right)^{1/2} g(n) \\
&= \left( \frac{n^2+n+1/4}{n^2+n} \right)^{1/2} g(n) \ > \ g(n) \ .
\end{aligned}
$$

Hence, $g(n)$ is a monotonic increasing sequence in $n$ and $g(n) \geq 1/2$. Therefore,

$$-\log(g(n)) \leq -\log(1/2) = 1$$

and the lemma holds for all $n \geq 1$. ∎

Based on the above lemma we now give our main competitiveness result.

**Theorem 2** Let $\mathbf{x}^1, \mathbf{x}^2, \ldots, \mathbf{x}^T (T \geq 2)$ be any sequence of price relatives for $N$ assets. Let $S_T(\mathcal{Q})$ be the wealth of a switching regime $\mathcal{Q}$ that uses pure investment strategies (as defined by Equation (3)) and let $l(\mathcal{Q})$ be the number of times $\mathcal{Q}$ switches from one strategy to another. Then, the logarithmic wealth achieved by the switching portfolios algorithm is at least

$$
\begin{aligned}
\log(S_T(\mathcal{Q})) \ &- \ \frac{3}{2}l(\mathcal{Q})\log\left( \frac{T}{l(\mathcal{Q})} \right) \\
&- \ 1/2\log(T) - (l(\mathcal{Q})+1)\log(4N) \ .
\end{aligned}
$$

**Proof** The wealth achieved by the switching portfolios algorithm as given by Equation (8) is an efficient way to calculate the sum $\sum_{\mathcal{Q}'} P_0(\mathcal{Q}')S_T(\mathcal{Q}')$. Thus, the logarithmic wealth achieved by the switching portfolios algorithm is at least

$$
\begin{aligned}
\log &\left( \sum_{\mathcal{Q}'} P_0(\mathcal{Q}')S_T(\mathcal{Q}') \right) \\
&\geq \ \log(P_0(\mathcal{Q})S(\mathcal{Q})) \\
&= \ \log(P_0(\mathcal{Q})) + \log(S(\mathcal{Q})) \ . \qquad (9)
\end{aligned}
$$

From the definition of $\hat{\gamma}(\Delta t)$ we obtain

$$
\begin{aligned}
P_0(\mathcal{Q}) \ = \ &\frac{1}{N}(N-1)^{-l} \\
&\left( \prod_{i=1}^{l} \left( \prod_{j=1}^{t_i-t_{i-1}-1} \frac{j-1/2}{j} \right) \frac{1/2}{t_i-t_{i-1}} \right) \\
&\prod_{j=1}^{T-t_l-1} \frac{j-1/2}{j} \ .
\end{aligned}
$$

Let $l$ be a shorthand for $l(\mathcal{Q})$. Using Lemma 1 and simple algebraic manipulations, we get that,

$$
\begin{aligned}
-\log(P_0(\mathcal{Q})) \ = \ &\log(N) + l\log(N-1) \\
&+ \ 3/2\sum_{i=1}^{l}\log(t_i - t_{i-1}) \\
&+ \ 2l + 1/2\log(T-t_l) + 1 \ .
\end{aligned}
$$

Using the log-sum inequality [9] we can bound the above expression as follows,

$$
\begin{aligned}
&-\log(P_0(\mathcal{Q})) \\
&\leq \ \log(N) + l\log(N-1) + 3/2l\log(T/l) \\
&\quad + 1/2\log(T) + 2l + 1 \\
&\leq \ (l+1)\log(4N) \\
&\quad + 3/2l\log(T/l) + 1/2\log(T) \ . \qquad (10)
\end{aligned}
$$

Finally, the theorem is proved combining Equation (9) with Equation (10). ∎

Deriving a competitiveness bound for the first version of the algorithm is a much easier task since $\gamma$ is fixed. In short, using the (fixed) value of $\gamma$ in Equation (9) for $P_0(\mathcal{Q})$, we get that logarithmic wealth achieved by the switching portfolios algorithm with fixed $\gamma$ compared to any switching regime is smaller by at most,

$$(l+1)\log(N) + l\log(1/\gamma) + (T-l)\log(1/(1-\gamma)) \ .$$

The competitiveness bound of Theorem 2 has a nice intuitive interpretation. Although a switching regime that frequently switches from one investment strategy to another has the potential of achieving high returns, the difference in the wealth achieved by the switching portfolios algorithm and a frequently switching regime is large. Furthermore, since the bound holds for any switching regime we in fact get that the wealth achieved by the algorithm is at least,

$$\max_{\mathcal{Q}} \left\{ \begin{array}{l} \log(S_T(\mathcal{Q})) - \frac{3}{2}l(\mathcal{Q})\log(\frac{T}{l(\mathcal{Q})}) \\ -\frac{1}{2}\log(T) - l(\mathcal{Q})\log(4N) \end{array} \right\} \ .$$

Therefore, the switching portfolios algorithm encompasses a natural tradeoff between the yield of a switching regime and its complexity (in terms of the number of time it switches).

In conclusion of this section, we would like to note that there are other prior distributions for switching models that lead to efficient algorithms in similar settings (see for instance [12]). The switching portfolio algorithm itself can be used with other probability distributions for switching sequences. For instance, in settings where the best model changes frequently a prior that favors frequent switches can be employed. Similar prior distributions that constrain the maximal time each asset can be held may also be used in order to reduced the computation time.

## 4   Transaction costs

Coping with transaction costs when using the switching portfolios algorithm with pure investment strategies is relatively a simple task. Since on each trading day we take a portion of wealth from each asset and redistribute it among the rest of the assets, we simply need to deduct the cost of selling asset $i$, and then further deduct the cost of buying asset $j$. Clearly, this is not the least expensive scheme to redistribute wealth. We might end up selling and buying the same asset and thus pay more commission than it would have been needed had we pre-calculated the amount we need to sell/buy per each asset. (We actually found and used the optimal trading procedure for each portfolio selection algorithm in the experiments reported later in this section.) However, it is enough to use this scheme in order to achieve the same competitiveness bounds. For the case of a fixed percentage $\mathbf{c}$ of the amount we trade (either buy or sell), the first equation of wealth update (Equation (7)) remains the same (no trading is taking place), while Equation (8) becomes now,

$$S_{t+1,t+1}^i =$$
$$(1-\mathbf{c})^2 \left( \sum_{j\neq i} \sum_{t_0=1}^{t} \frac{1}{2(t-t_0+1)} S_{t,t_0}^j \right) x_i^{t+1} \ .$$

Although simplistic, this approach results in the same competitiveness bounds when using only pure strategies.

When we switch from one strategy to another we have to sell entirely one asset and buy a new one. Therefore, the wealth achieved by a switching regime $\mathcal{Q}$ in the presence of fixed percentage transaction costs is now

$$S_T(\mathcal{Q}) = \prod_{j=1}^{l+1} (1-\mathbf{c})^2 \prod_{t=t_{j-1}+1}^{t_j} x_{i_j}^t \ .$$

Thus, the bound on the difference between the wealth achieved by the switching portfolios algorithm and any specific switching regime does not change.

It is also simple to derive a wealth update scheme for other transaction costs models. For instance, Blum and Kalai (1997) performed experiments with a transaction costs model where first the entire portfolio vector is updated in *parallel* for all the assets, then the cost is subtracted from the total wealth (some from each stock) such that the wealth proportions would not change. Formally, the transaction cost of changing the wealth distribution from $\{S_i\}$ to $\{S_i'\}$ in this model is, $c\sum_i |S_i - S_i'|$. The analysis for this transaction cost model follows similar lines. For a switching regime $\mathcal{Q}$, we either keep the same asset or sell it completely and buy a new asset. Thus, the former case incur a zero transaction cost while the latter decreases wealth by a factor of $1-2\mathbf{c}$. Therefore, the amount invested in the $i$th asset in case of a switch, previously given by Equation (8), is now calculated as follows

$$S_{t+1,t+1}^i =$$
$$(1-2\mathbf{c}) \left( \sum_{j\neq i} \sum_{t_0=1}^{t} \frac{1}{2(t-t_0+1)} S_{t,t_0}^j \right) x_i^{t+1} \ .$$

Again, the bound on wealth achieved by the switching portfolios algorithm remains the same.

As mentioned before, Blum and Kalai used the above transaction cost model in their experiments with their generalization of Cover's universal portfolio algorithm. In Table 2 we compare the wealths achieved by the generalized universal portfolio algorithm and the switching portfolios algorithm for the same subsets of stocks appearing in Table 1 using the 'parallel transaction costs' model. The results are given for transaction costs of 2% and 5%. In all the cases we checked with these transaction costs, the switching portfolios algorithm achieves better results than the generalized universal portfolio algorithm. Furthermore, in several cases the switching portfolios algorithm also outperforms the best CRP.

We also checked the performance on larger sets of stocks that were not reported in previous papers. In all cases checked, we found that the returns of the universal portfolio algorithm were smaller than those of the switching portfolios algorithm, and often smaller than the final closing value of the best constituent stock in the portfolio. Moreover, we found that the universal portfolio scales rather poorly with the size of the portfolio, i.e., the number of stocks, as indeed the competitiveness bound for the universal portfolio algorithm implies. The performance of the switching portfolios algorithm also degrades as the size of the portfolio gets larger, however, we found that in practice the degradation is less severe.

| Stocks | 2% Transaction Costs | | | 5% Transaction Costs | | |
|---|---|---|---|---|---|---|
| | BCRP | Universal | Switching | BCRP | Universal | Switching |
| Iroquois & Kin Ark | 22.06 | 17.05 | 22.89 | 8.92 | 4.03 | 15.55 |
| Com. Met. & Kin Ark | 59.07 | 36.57 | 42.52 | 52.02 | 11.00 | 33.66 |
| Com. Met. & Mei. Corp. | 52.98 | 43.04 | 67.60 | 52.02 | 18.29 | 64.50 |
| IBM & Coca-Cola | 13.36 | 11.33 | 13.82 | 13.36 | 7.97 | 13.38 |

Table 2: Comparison of wealths achieved by the universal portfolio algorithm and the switching portfolios algorithm with 2% and 5% transaction costs. For comparison the wealth achieved by the best CRP is also provided.

To conclude this section, we give an illustrative example of the behavior of the switching portfolios algorithm. In all figures mentioned below which describe wealth growth, the $x$ axis corresponds to the number of trading days since January 1963 and the $y$ axis is the wealth achieved by the various algorithms assuming an initial investment of one unit before the first trading day (January 1st 1963). In the example we used three stocks: Dow Chemicals, Espey Manufacturing, and Kin Ark. These stocks are rather volatile and, as shown at the top part of Figure 3, they exhibit different behavior during relatively long periods. In Figures 1 and 2 we show the wealths achieved by the switching portfolios algorithm and the generalized universal portfolio, on each of the trading days, for transaction costs of 0.5% and 3%, respectively. For comparison, we also give the daily price of the best constituent stock. In both cases the switching portfolio algorithm clearly outperforms the universal portfolio algorithm. Furthermore, in the presence of a hefty 3% commission for selling and buying stocks, one is better of with keeping a single stock rather than using the universal portfolio algorithm. The wealth achieved by the switching portfolio algorithm also decreases with the increase in the commission. However, the algorithm still outperforms any single stock even in the presence of high transaction costs. Last, the bottom part of Figure 3 illustrates the behavior of the switching portfolios algorithm on this portfolio containing three stocks (for a 3% transaction cost). In the figure we used different gray levels to indicate the largest asset held by the algorithm on each trading day. That is, if $\forall j \neq i : S_t^i > S_t^j$ then we draw a thin line using the $i$th gray level at the location corresponding to the $t$th trading day. One can see from the figure that there are long periods where a significant portion of wealth is invested in a single asset, which is the stock that *locally* achieves the highest returns. This behavior confirms empirically our basic assumption that high returns can be achieved by using a switching regime that infrequently alternates between the possible assets.

It is also possible to use more complex investment strategies, in addition to (or instead of) pure strategies, as the set of basic investment strategies. For instance, Blum and Kalai's universal portfolio algorithm itself can be used as a basic investment strategy and 'fed' into the switching portfolios algorithm. We implemented and evaluated a version of the switching portfolios algorithm that includes the generalized universal portfolio algorithm as a basic strategy (in addition to the pure strategies). The time complexity of this version grows like $O(t^3)$ since at the $t$th trading day we need to maintain $t$ different copies of universal portfolios, one for each possible starting date $1 \leq t_0 \leq t$. In the experiments we performed, we found only a modest improvement over using pure strategies alone as our basic investment strategies. These findings are not surprising since, as argued earlier, with the benefit of hindsight, we can use pure investment and still gain enormous wealth. Put another way, the switching portfolios algorithm tries to imitate a prescient observer and thus the improvement using more complex investment strategies, that assume the stock market is stationary, is relatively small.

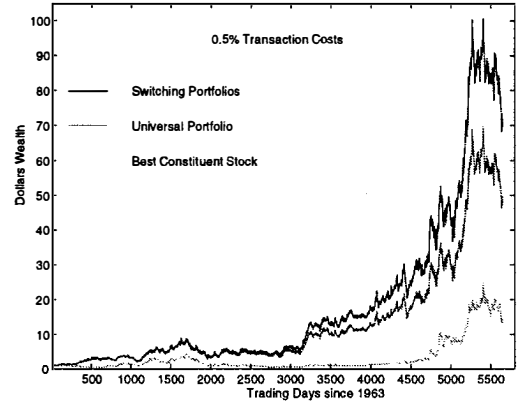

Figure 1: Daily wealths achieved by the switching portfolios algorithm (with pure strategies), the generalized universal portfolio algorithm, and the daily price of best constituent stock in the presence of 0.5% transaction costs. The stocks constituting the portfolio are: Dow Chemicals, Espey Manufacturing, and KinArk.

## 5  Conclusions

A simple and efficient portfolio selection algorithm was presented in this paper. The algorithm is competitive with any switching regime determined in hindsight. Similar approaches were investigated for different settings [12, 10]. One of the contributions of this paper is the distillation of key elements of previously known methods, and the synthesis with other learning-theoretic and information-theoretic results that lead to an algorithm that outperforms previously published rebalancing algorithms for portfolio selection. Furthermore, our algorithm achieves a slightly better competitiveness bound than the bounds reported in [12, 10]. In addition, it also can use *non-fixed* investment strategies as discussed above. The price that we pay for the improved bounds and flexibility is a more complex and time con-suming algorithm. A challenging question is whether our algorithm can be implemented using a constant time per asset for each trading day.

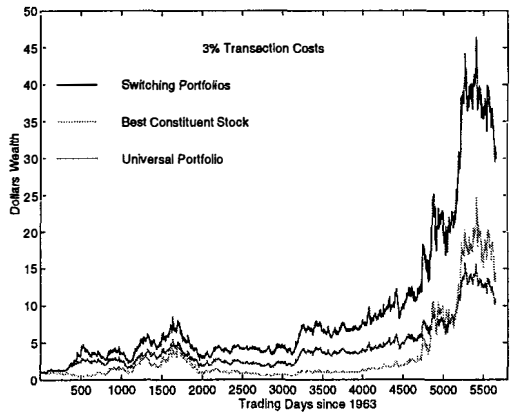

Figure 2: Daily wealths achieved by the switching portfolios algorithm (with pure strategies), the generalized universal portfolio algorithm, and the price value of best constituent stock in the presence of 3% transaction costs for the same stocks as in Figure 1.

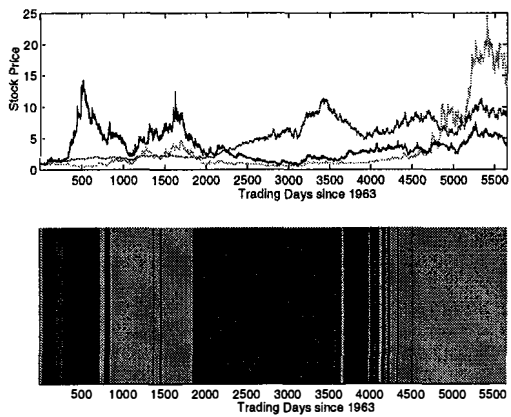

Figure 3: Top: the daily price of Kin Ark (dark gray), Dow Chemicals (medium gray), and Espey Manufacturing (light gray). Bottom: gray level encoding description of the largest asset among the above stocks held by the switching portfolios algorithm on each trading day.

There is still some room for improvement within the current framework. First, we use the least informative scheme, namely the uniform distribution, to redistribute wealth upon a switch. More informative wealth redistribution techniques that take into account past performance might yield higher returns as well as better competitiveness bounds. Second, our switching model employs a tacit assumption that the *prior* probability of switching is fixed. However, real stock markets alternate between periods of rapid price changes and relatively calm periods. Such behavior is not taken into account by the current *prior* probability model for switching. It might be possible to model time dependent prior distributions using hyper-parameterization for the switching probability. Last, although we observed only a modest improvement in performance when we used complex investment strategies, this might simply reflect a limitation of the universal portfolio algorithm. Investigating alternative basic investment strategies that may result in higher returns is one of our future research goals.

The switching portfolios algorithm still overlooks several important issues. The switching portfolios algorithm, as well as the other portfolio rebalancing techniques mentioned in this paper, ignore any risk and volatility factors that might be crucial for individual investors. Second, when trading in large volumes, wealth redistributions might cause a significant and instantaneous change in the price of the assets that were traded. Thus, a model that takes into account self-influence due to trading on stock prices would be necessary for large volume portfolio managers. Lastly, there are several external factors and sources of informations that are neglected. Some of these factors can be dealt within the current framework using the simple model of "side information" (see [11, 8]). However, most factors, such as taxation, which largely influence the actual yield cannot be incorporated using the "side information" model.

**Acknowledgments**  Thanks to Yoshua Bengio, Mark Herbster, Rob Schapire, Manfred Warmuth, and Andreas Weigend for helpful discussions.